\documentclass[12pt,english]{article}
\usepackage[T1]{fontenc}
\usepackage[latin9]{inputenc}
\usepackage[letterpaper]{geometry}
\usepackage{float}
\usepackage{units}
\usepackage{textcomp}
\usepackage{url}
\usepackage{amsthm}
\usepackage{amsmath}
\usepackage{amssymb}

\makeatletter

\providecommand{\tabularnewline}{\\}
\floatstyle{ruled}
\newfloat{algorithm}{tbp}{loa}
\floatname{algorithm}{Algorithm}

 \theoremstyle{definition}
 \newtheorem*{defn*}{Definition}
  \theoremstyle{remark}
  \newtheorem*{rem*}{Remark}

\usepackage{amsmath}
\usepackage{amssymb}
\usepackage{url}
\usepackage{listings}
\usepackage{url}

\makeatother

\usepackage{babel}

\begin{document}

\title{Series misdemeanors}

\author{David R. Stoutemyer%
\thanks{dstout at hawaii dot edu%
}}
\maketitle
\begin{abstract}
Puiseux series are power series in which the exponents can be fractional
and/or negative rational numbers. Several computer algebra systems
have one or more built-in or loadable functions for computing truncated
Puiseux series -- perhaps generalized to allow coefficients containing
functions of the series variable that are dominated by any power of
that variable, such as logarithms and nested logarithms of the series
variable. Some computer-algebra systems also offer functions that
can compute more-general truncated recursive \textsl{hierarchical}
series. However, for all of these kinds of truncated series there
are important implementation details that haven't been addressed before
in the published literature and in current implementations.

For implementers this article contains ideas for designing more convenient,
correct, and efficient implementations or improving existing ones.
For users, this article is a warning about some of these limitations.
Many of the ideas in this article have been implemented in the computer-algebra
within the TI-Nspire calculator, Windows and Macintosh products. 
\end{abstract}

\section{Introduction}

Here is a conversation recently overheard at a car-rental desk: 
\begin{quote}
\textsl{Customer:} {}``I followed your directions of three right
turns to get on the highway, but that put me in a fenced corner from
which I could only turn right, bringing me back to where I started!''

\textsl{Agent}: {}``Make your first right turn \textsl{after} exiting
the rental car lot.'' 
\end{quote}
The original directions were correct, but incomplete.

The same was true of published algorithms for truncated Puiseux series.
After reading all that I could find about such algorithms, I implemented
them for the computer algebra embedded in the TI Nspire\texttrademark{}
handheld graphing calculator, which also runs on PC and Macintosh
computers. Testing revealed some incorrect results due to ignorance
about some important issues. Results for other series implementations,
reveal that their implementers have made similar oversights.

It required a significant effort to determine how to overcome these
difficulties. This article is intended as a warning for users of implementations
that exhibit the flaws -- and as suggestions to implementers for repairing
those flaws or avoiding them in new implementations. Many of the ideas
here are implemented in TI-Nspire.

Additional issues for truncated and infinite series are described
in a sequel to this article tentatively titled {}``Series crimes''
\cite{StoutemyerSeriesCrimes}.

For real-world problems, exact closed-form symbolic solutions are
less frequently obtainable than are various symbolic series solutions.
Therefore in practice, symbolic series are among the most important
features of computer algebra systems.

Almost all computer algebra systems have a function that produces
at least truncated Taylor series. Iterated differentiation followed
by substitution of the expansion point provides a very compact implementation.
However, it can consume time and data memory that grows painfully
with the requested order of the result. Knuth \cite{Knuth} presents
algorithms that are significantly more efficient for addition, multiplication,
raising to a numeric power, exponentials, logarithms, composition
and reversion. He also suggests how to derive analogous algorithms
for any function that satisfies a linear differential equation. Silver
and Sullivan \cite{SilverAndSullivan} additionally give such algorithms
for sinusoids and hyperbolic functions. Brent and Kung \cite{BrentAndKung}
pioneered algorithms that are faster when many non-zero terms are
needed.

If for a truncated Puiseux series expanded about $z=0$ the degree
of the lowest-degree non-zero term is $\alpha$ and $g$ is the greatest
common divisor of the increments between the exponents of non-zero
terms, then the mapping $z^{\beta}\rightarrow t^{\left(\beta-\alpha\right)/g}$
can be used, with care, to adapt many of the Taylor series algorithms
for Puiseux series.

As described by Zipple \cite{ZippleConference,ZippleBook}, many Puiseux
series implementations have generalized them to allow coefficients
that contain appropriate logarithms and nested logarithms that depend
on the series variable. Geddes and Gonnet \cite{GeddesAndGonnet},
Gruntz \cite{Gruntz}, and Richardson \textit{et. al.} \cite{RichardsonEtAl}
give algorithms for more general truncated \textsl{hierarchical} series
that also correctly prioritize essential singularities and perhaps-nested
logarithms in coefficients. Koepf \cite{Koepf} implemented \textsl{infinite}
Puiseux series in which the result is expressed as a symbolic sum
of terms: The general term in the summand typically depends on the
summation index and powers of the series variable. Some implementations
compute Dirichlet, Fourier, or Poisson series. Most of the issues
described in this article are relevant to most kinds of truncated
series, and some of the issues are also relevant to infinite series.

To obtain a series for $f\left(w\right)$ expanded about $w=w_{0}$
with $w_{0}$ finite and non-zero, we can substitute $w\rightarrow z+w_{0}$
into $f\left(w\right)$ giving $g\left(z\right),$ then determine
the series expansion of $g\left(z\right)$ about $z=0,$ then back
substitute $z\rightarrow w-w_{0}$ into that result. However, if the
series is to be used only for real $w<w_{0}$ then using instead $w\rightarrow w_{0}-z$
might give a result that more candidly avoids unnecessary appearances
of $i$ -- particularly if $f\left(w\right)$ contains logarithms
or fractional powers.

To obtain a series for $f\left(w\right)$ expanded from the complex
circle of radius $\infty,$ we can substitute $\zeta\rightarrow1/z$
into $f\left(w\right)$ giving $h\left(z\right)$, then determine
the series expansion of $h\left(z\right)$ about $z=0,$ then back
substitute $z\rightarrow1/w$ into that result.

A proper subset of the complex circle at infinity can be expressed
by an appropriate constraint on the series variable, such as\begin{eqnarray*}
\mbox{series}\left(f\left(w\right),w=-\infty,\ldots\right) & \rightarrow & H\left(\dfrac{1}{w}\right)\end{eqnarray*}
 where\begin{eqnarray*}
H\left(x\right) & = & \mbox{series}\left(f\left(\dfrac{1}{x}\right),x=0,\ldots\right)\:|\: x<0.\end{eqnarray*}
 Therefore without loss of generality, the expansion point is $z=0$
throughout the remainder of this article with, $z=x+iy=re^{i\theta}$
where $r\geq0$, $-\pi<\theta\leq\pi$, and $x,y\in\mathbb{R}.$

Also, wherever braced case constructs occur, the tests are presumed
to be done using short-circuit evaluation from top to bottom to avoid
the clutter of making the tests mutually exclusive.

\section{The disorder of order}

\begin{flushright}
\textsl{{}``What we imagine is order is merely the prevailing form
of chaos.}''\\
 -- Kerry Thornley 
\par\end{flushright}

Truncated series functions usually have a parameter by which the user
requests a certain numeric {}``order'' for the result. Existing
implementations treat this request in different ways, some of which
are significantly more useful than others.

\subsection{Render onto users what they request}

\begin{flushright}
\textsl{{}``Good order is the foundation of all things.''}\\
 -- Edmund Burke 
\par\end{flushright}

\begin{flushright}
\textsl{{}``Mathematics is the art of giving the same name to different
things.''}\\
 -- Jules Henri Poincaré 
\par\end{flushright}

Unfortunately, the word {}``order'' is used in too many ways in
mathematics. Relevant definitions used in this article are:
\begin{defn*}
The \textsl{exact error order} of a truncated series result expanded
about $z=0$ is $\tau$ if the error is $O\left(z^{\tau}\right)$
but the error isn't $o\left(z^{\tau}\right)$. The exact error order
of an exact series result expanded about $z=0$ is $+\infty$.\end{defn*}
\begin{rem*}
Knuth \cite{KnuthBigTheta} introduced the convenient notation $\Theta\left(z^{\tau}\right)$
to denote exact-order $\tau$ in $z$. In comparison to $O\left(z^{\tau}\right)$,
$\Theta\left(z^{\tau}\right)$ avoids discarding valuable information
when we also know that a result isn't $o\left(z^{\tau}\right)$.\end{rem*}
\begin{defn*}
The \textsl{degree of a truncated Puiseux series} with respect to
$z$ expanded about $z=0$ is the largest exponent of $z$ that occurs
outside of any argument of any $O\left(\ldots\right)$, $o\left(\ldots\right)$,
or $\Theta\left(\ldots\right)$ term that is included in the result. 
\end{defn*}
It is unreasonable to request a \textsl{degree} because, for example,
there is no way for\[
\textrm{series}\left(\cos z,\, z\!=\!0,\mathrm{\, degree}\!=\!1\right)\]
 to return a result of degree 1. It is also unreasonable to request
an \textsl{exact error order} because, for example, $\textrm{series}\left(\cos z,\, z\!=\!0,\,\Theta\left(z^{3}\right)\right)$
can't return a series having error $\Theta\left(z^{3}\right)$.
\begin{defn*}
If a series-function order-argument $\tau$ denotes a request that
the result be $\cdots+O\left(z^{\tau}\right)$, then the degree of
an \textsl{as-requested big-O} result should be the largest degree
that satisfies\begin{equation}
\mathrm{degree}<\tau\leq\mathrm{exact\: error\: order}.\label{eq:AsRequestedBigO}\end{equation}
 \end{defn*}
\begin{rem*}
It seems likely that more often users prefer to specify the highest
degree term they \textsl{want} to view rather than the lowest degree
term they \textsl{don't want} to view. Thus most users would prefer
that the series function parameter $\tau$ denotes a request for a
result that is $\cdots+o\left(z^{\tau}\right)$. Therefore:\end{rem*}
\begin{defn*}
If a series-function order-argument $\tau$ denotes a request that
the result be $\cdots+o\left(z^{\tau}\right)$, then the degree of
an \textsl{as-requested little-o} result should be the largest degree
that satisfies\begin{equation}
\mathrm{degree}\leq\tau<\mathrm{exact\: error\: order}.\label{eq:AsRequestedLittleo}\end{equation}

\end{defn*}
The Maxima, Mathematica$^{\text{\textregistered}}$ and TI-Nspire\texttrademark{}
truncated Puiseux series functions use this little-$o$ interpretation
of a numeric parameter $\tau$. For example, glossing over their input
and output syntax differences, they all give\begin{eqnarray}
\mathrm{series}\left(\dfrac{\sin z}{z^{3}},\, z\!=\!0,\,5\right) & \rightarrow & \dfrac{1}{z^{2}}-\dfrac{1}{6}+\dfrac{z^{2}}{120}-\dfrac{z^{4}}{5040}.\label{eq:SeriesSinzOnzCubed}\end{eqnarray}
 To this Maxima appends {}``$+\cdots$'' and Mathematica appends
{}``$+\, O[z]^{6\,}$''.

It is easier to implement an interpretation in which the order parameter
$\tau$ denotes that the inner-most sub-expressions are computed to
$o\left(z^{\tau}\right)$ and the final result is computed to whatever
order that yields. For reasons described below, that can and often
does lead to a result that is $o\left(z^{\kappa}\right)$ with a $\kappa$
that is smaller or occasionally larger than $\tau$. For example,
it would omit the last term of result (\ref{eq:SeriesSinzOnzCubed}).
For such an implementation it is essential to display an error term
because otherwise:\vspace{-0.1in}

\begin{enumerate}
\item If exact order $\leq$ requested little-$o$ order, then the result
doesn't reveal that it is less accurate than requested, which can
be disastrous.\vspace{-0.1in}

\item If degree > requested little-$o$ order, then the user must notice
and perhaps somehow truncate the excess terms to use the result in
further calculations as intended. 
\end{enumerate}
However, even if there is an error term indicating a result that doesn't
have the requested accuracy, this design is inconvenient for users
because:\vspace{-0.1in}

\begin{enumerate}
\item Users often don't notice the deficient or excessive accuracy.\vspace{-0.1in}

\item Users who notice excessive order, must perhaps somehow truncate the
excess terms to use the result in further calculations as intended.\vspace{-0.1in}

\item Users who notice deficient order are forced to iteratively guess the
order argument to use in series$\left(\ldots\right)$ to obtain sufficient
accuracy -- then perhaps somehow truncate a result that exceeds the
desired order.\vspace{-0.1in}

\item If the user is another function, then that function should test the
returned order and correct it if necessary by iterative adjustment
and/or truncation. This is a requirement that might not occur to many
authors of such functions -- particularly those who aren't professional
computer-algebra implementers. 
\end{enumerate}
It is more considerate, reliable and efficient to build any necessary
iterative adjustment and/or truncation into the series$\left(\ldots\right)$
function rather than to foist it on all function implementers and
top-level users. It isn't prohibitively harder to implement an as-requested
result.

\subsection{How to deliver as-requested order\label{sub:How-to-deliver}}
\begin{defn*}
If an infinite Puiseux series is 0, then its \textsl{dominant term
}is 0. Otherwise the dominant term is the lowest-degree non-zero term.\vspace{-0.3in}

\end{defn*}
$\,$ 
\begin{defn*}
If an infinite Puiseux series is 0, then its\textsl{ dominant exponent
}is $\infty$. Otherwise the dominant exponent is the exponent of
the dominant term.\vspace{-0.2in}

\end{defn*}
$\,$
\begin{rem*}
Some authors call the dominant exponent the \textsl{valuation} or
\textsl{valence}, but other authors confusingly call it the \textsl{order}.
\end{rem*}
A typical truncated-Puiseux-series implementation recursively computes
series for the operands of each operator and the arguments of each
function, combining those series according to various algorithms.
Table \ref{Flo:TableNecessaryOperandOrders}, lists the dominant exponent
of a result and the operand orders that are necessary and sufficient
to determine a result to $o\left(z^{k}\right)$. In that table a result
dominant exponent of $-\infty$ signifies an essential singularity.

\begin{table}[ht]
\caption{\label{tab:necessaryOperandOrders}$\protect\begin{array}{cc}
\mathrm{Requested\: operand\: orders\: for\: a\: result\: having\: order\:}o\left(z^{k}\right),\:\mathrm{with}\protect\\
U\!=cz^{\alpha}\!+\! bz^{\sigma}\!+\!\cdots\!+\! o\left(z^{m}\right)\:\mathrm{and}\: V\!=az^{\beta}\!+\! hz^{\gamma}\!+\!\cdots\!+\! o\left(z^{n}\right)\protect\\
\mathrm{where}\:\alpha,\beta,\sigma,\gamma,m,n,k\in\mathbb{Q}\protect\end{array}$}

\label{Flo:TableNecessaryOperandOrders}\begin{tabular}{|c|c|c|}
\hline 
operation  & result dominant exponent  & request $m$ and $n$\tabularnewline
\hline
\hline 
$U\pm V$  & $\mathrm{\geq min}\left(\alpha,\beta\right)$  & $\begin{array}{c}
m=n=k\end{array}$\tabularnewline
\hline 
$UV$  & $\geq\alpha+\beta$  & $\begin{array}{c}
m=k-\beta\\
n=k-\alpha\end{array}$\tabularnewline
\hline 
$\dfrac{U}{V}$  & $\geq\alpha-\beta$  & $\begin{array}{c}
m=k+\beta\\
n=k-\alpha+2\beta\end{array}$\tabularnewline
\hline 
$U^{\gamma}$  & $\geq\gamma\alpha$  & $m=k+(1-\gamma)\alpha$\tabularnewline
\hline 
$\begin{array}{c}
e^{U}\\
\cos U\\
\cosh U\end{array}$  & $\begin{cases}
0 & \mathrm{if}\:\alpha\geq0\\
\mathrm{-\infty} & \mathrm{otherwise}\end{cases}$  & $m=\begin{cases}
k & \mathrm{if}\:\alpha\geq0\\
\mathrm{essential\: singularity} & \mathrm{otherwise}\end{cases}$\tabularnewline
\hline 
$\begin{array}{c}
\sin U\\
\tan U\\
\sinh U\\
\tanh U\end{array}$  & $\begin{cases}
\alpha & \mathrm{if}\:\alpha\geq0\\
-\infty & \mathrm{otherwise}\end{cases}$  & $m=\begin{cases}
k & \mathrm{if}\:\alpha\geq0\\
\mathrm{essential\: singularity} & \mathrm{otherwise}\end{cases}$\tabularnewline
\hline 
$\ln U$  & $\begin{cases}
\sigma & \mathrm{if}\: cz^{\alpha}=1\\
0 & \mathrm{otherwise}\end{cases}$  & $m=k+\alpha$\tabularnewline
\hline 
$\mathrm{arctanh}\, U$  & $\begin{cases}
\alpha & \mathrm{if}\:\alpha\geq0\\
0 & \mathrm{otherwise}\end{cases}$  & $m=\begin{cases}
k+\sigma & \mathrm{if}\: cz^{\alpha}=1\vee cz^{\alpha}=-1\\
k+2\alpha & \mathrm{if}\:\alpha<0\\
k & \mathrm{otherwise}\end{cases}$\tabularnewline
\hline 
$\arctan U$  & $\begin{cases}
\alpha & \mathrm{if}\:\alpha\geq0\\
0 & \mathrm{otherwise}\end{cases}$  & $m=\begin{cases}
k+\sigma & \mathrm{if}\: cz^{\alpha}=i\vee cz^{\alpha}=-i\\
k+2\alpha & \mathrm{if}\:\alpha<0\\
k & \mathrm{otherwise}\end{cases}$\tabularnewline
\hline 
$\mathrm{arcsinh}\, U$  & $\begin{cases}
\alpha & \mathrm{if}\:\alpha\geq0\\
0 & \mathrm{otherwise}\end{cases}$  & $m=\begin{cases}
k+\sigma/2 & \mathrm{if}\: cz^{\alpha}=i\vee cz^{\alpha}=-i\\
k+\alpha, & \alpha<0\\
k & \mathrm{otherwise}\end{cases}$\tabularnewline
\hline 
$\arcsin U$  & $\begin{cases}
\alpha & \mathrm{if}\:\alpha\geq0\\
0 & \mathrm{otherwise}\end{cases}$  & $m=\begin{cases}
k+\sigma/2 & \mathrm{if}\: cz^{\alpha}=1\vee cz^{\alpha}=-1\\
k+\alpha, & \alpha<0\\
k & \mathrm{otherwise}\end{cases}$\tabularnewline
\hline 
$\begin{array}{c}
\arccos U\\
\mathrm{arccosh}\, U\end{array}$  & $\begin{cases}
\sigma/2 & \mathrm{if}\: cz^{\alpha}=1\\
0 & \mathrm{otherwise}\end{cases}$  & $m=\begin{cases}
k+\sigma/2 & \mathrm{if}\: cz^{\alpha}=1\vee cz^{\alpha}=-1\\
k+\alpha & \mathrm{if}\:\alpha<0\\
k & \mathrm{otherwise}\end{cases}$\tabularnewline
\hline
\end{tabular}
\end{table}

As indicated there, cancellation of the dominant terms of series $U$
and $V$ can cause the dominant exponent of $U\pm V$ to exceed $\mathrm{min}\left(\alpha,\beta\right)$
when $\alpha=\beta$, such as for\begin{eqnarray*}
\mathrm{series}\left(e^{z}-\cos z,\, z\!=\!0,\, o\left(z\right)\right) & \rightarrow & \left((1+z+o(z)\right)-\left(1+o(z)\right)\\
 & \rightarrow & z+o(z).\end{eqnarray*}

If the coefficient domain has zero-divisors, such as for modular arithmetic
or floating-point with underflow, then the dominant exponent of $UV$
can exceed $\alpha+\beta$, and the dominant exponent of $U^{\gamma}$
can exceed $\gamma\alpha$.

Unfortunately, most of the entries in column 3 require us to know
the dominant exponents of the operands, perhaps also together with
a dominant coefficient $c$ and the exponent $\sigma$ of the next
non-zero term, if any. Therefore we need this information \textsl{before}
computing the operand series to the correct order, but we don't have
this information until \textsl{after} we have computed the first term
or two of the operand series.

One way to overcome the difficulty is as follows: We can \textsl{guess}
the dominant exponent of the operands by using a function written
according to rewrite rules such as the following, which are heuristically
motivated by the second column of Table \ref{tab:necessaryOperandOrders}:\begin{eqnarray*}
\mbox{guessDE}\left(z,z\right) & \rightarrow & 1,\\
\mbox{guessDE}\left(u+v,z\right) & \rightarrow & \min\left(\mbox{guessDE}\left(u,z\right),\mbox{guessDE}\left(v,z\right)\right),\\
\mbox{guessDE}\left(uv,z\right) & \rightarrow & \mathrm{guessDE}\left(u,z\right)+\mathrm{guessDE}(v,z),\\
\mbox{guessDE}\left(u^{k},z\right) & \rightarrow & k\,\mbox{guessDE}\left(u,z\right),\\
\mbox{guessDE}\left(e^{u},z\right) & \rightarrow & 0,\\
\mbox{guessDE}\left(\sin u,z\right) & \rightarrow & \max\left(0,\mbox{guessDE}\left(u,z\right)\right),\\
\mbox{guessDE}\left(\ln u,z\right) & \rightarrow & \begin{cases}
\mbox{guessDE}\left(u-1,z\right), & \mathrm{if\:}u(0)=1,\\
0, & \mathrm{otherwise},\end{cases}\\
\mbox{guessDE}\left(\arctan u,z\right) & \rightarrow & \begin{cases}
\mbox{guessDE}\left(u-i,z\right), & \mathrm{if\:}u(0)=i,\\
\mbox{guessDE}\left(u+i,z\right), & \mathrm{if\:}u(0)=-i,\\
\max\left(0,\mbox{guessDE}\left(u,z\right)\right), & \mathrm{otherwise,}\end{cases}\\
\mbox{guessDE}\left(u,z\right)\;|\; u\:\mbox{is independent}\:\mbox{of}\: z & \rightarrow & 0.\end{eqnarray*}

If using the guess results in computing more terms than necessary,
then we should truncate the excess. If using the guess doesn't produce
the required order but reveals the dominant term (and where needed
the next non-zero term), then we know precisely the necessary and
sufficient order to request for recomputing the series operands.

If using the guess doesn't reveal this information, then when there
is only one function argument we can iteratively increase the guess,
starting with an initial increment $\delta>0$. For each iteration
we can double the increment added to the initial guess. This way,
in a modest multiple of the time required for the last iteration,
the process terminates successfully or by resource exhaustion.

Resource exhaustion can be caused by an undetected essential singularity,
insufficient simplification of an operand expression, or undetected
constancy around the expansion point, such as for $\left|x+1\right|+\left|x-1\right|$
at $x=0$, which can be more candidly expressed as\[
\begin{cases}
-2x & x<-1,\\
1 & -1\leq x\leq1,\\
2x & \mathrm{otherwise}.\end{cases}\]

To increase the likelihood of the first increment $\delta$ being
sufficient to expose the first non-zero term or two but not prohibitively
more terms than needed, we can use a function that guesses the increment
between the exponents of the first two non-zero terms. Let $u$ and
$v$ be expressions with $\tilde{\alpha}=\mathrm{guessDE}(u,z)$ and
$\tilde{\beta}=\mathrm{guessDE}(v,z)$. Then the following ordered
rewrite rules are examples for such a function:\begin{eqnarray*}
\textrm{guessInc}\left(z,z\right) & \rightarrow & 0,\\
\textrm{guessInc}\left(u^{\gamma},z\right) & \rightarrow & \textrm{guessInc}\left(u,z\right),\\
\textrm{guessInc}\left(uv,z\right) & \rightarrow & \begin{cases}
\mbox{guessInc}\left(u,z\right) & \mbox{guessInc}\left(v,z\right)=0,\\
\mbox{guessInc}\left(v,z\right) & \mbox{guessInc}\left(u,z\right)=0,\\
\mathrm{min}\left(\mbox{guessInc}\left(u,z\right),\mbox{guessInc}\left(v,z\right)\right), & \mathrm{otherwise},\end{cases}\\
\textrm{guessInc}\left(\ln u\right) & \rightarrow & \begin{cases}
\textrm{guessInc}\left(u-1,z\right) & \mathrm{if\:}u\left(0\right)=1,\\
\textrm{guessInc}\left(u,z\right) & \mathrm{otherwise},\end{cases}\\
\textrm{guessInc}\left(e^{u},z\right) & \rightarrow & \begin{cases}
\textrm{guessInc}\left(u,z\right) & \mathrm{if\:}\tilde{\alpha}=0,\\
\left|\tilde{\alpha}\right| & \mathrm{otherwise},\end{cases}\\
\textrm{guessInc}\left(\sin u,z\right) & \rightarrow & \begin{cases}
2\left|\tilde{\alpha}\right| & \mathrm{if\:}\textrm{guessInc}\left(u,z\right)=0,\\
\textrm{guessInc}\left(u,z\right) & \mathrm{otherwise},\end{cases}\quad//\:\mathrm{Same\: for\: sinh}\, u\\
\textrm{guessInc}\left(\cos u,z\right) & \rightarrow & \begin{cases}
\textrm{guessInc}\left(u,z\right) & \mathrm{if\:}\tilde{\alpha}=0,\\
2\left|\tilde{\alpha}\right| & \mathrm{otherwise},\end{cases}\quad//\:\mathrm{Same\: for\: cosh}\, u\\
\textrm{guessInc}\left(\arctan u,z\right) & \rightarrow & \begin{cases}
-\tilde{\alpha} & \mathrm{if}\:\tilde{\alpha}<0,\\
2\tilde{\alpha} & \mathrm{if\:\tilde{\alpha}>0\,\wedge\,}\textrm{guessInc}\left(u,z\right)=0,\\
\textrm{guessInc}\left(u-u(0),z\right) & \mathrm{if}\: u(0)=i\,\vee\, u(0)=-i,\\
\textrm{guessInc}\left(u,z\right) & \mathrm{otherwise},\end{cases}\end{eqnarray*}
 \begin{eqnarray*}
\textrm{guessInc}\left(\mathrm{arctanh}\, u,z\right) & \rightarrow & \begin{cases}
-\alpha & \mathrm{if}\:\alpha<0,\\
2\tilde{\alpha} & \mathrm{if\:\alpha>0\,\wedge\,}\textrm{guessInc}\left(u,z\right)=0,\\
\textrm{guessInc}\left(u-u(0),z\right) & \mathrm{if}\: u(0)=1\,\vee\, u(0)=-1,\\
\textrm{guessInc}\left(u,z\right) & \mathrm{otherwise},\end{cases}\\
\textrm{guessInc}\left(\arcsin u,z\right) & \rightarrow & \begin{cases}
2\left|\tilde{\alpha}\right| & \mathrm{if\:}\textrm{guessInc}\left(u,z\right)=0,\\
\textrm{guessInc}\left(u,z\right)/2 & \mathrm{if\:}u\left(0\right)=1\,\vee\, u\left(0\right)=-1,\\
\textrm{guessInc}\left(u,z\right) & \mathrm{otherwise},\end{cases}\\
\textrm{guessInc}\left(\mathrm{arcsinh}\, u,z\right) & \rightarrow & \begin{cases}
2\left|\tilde{\alpha}\right| & \mathrm{if\:}\textrm{guessInc}\left(u,z\right)=0,\\
\textrm{guessInc}\left(u,z\right)/2 & \mathrm{if\:}u\left(0\right)=i\,\vee\, u\left(0\right)=-i,\\
\textrm{guessInc}\left(u,z\right) & \mathrm{otherwise},\end{cases}\\
\textrm{guessInc}\left(\mathrm{arccosh}\, u,z\right) & \rightarrow & \begin{cases}
\tilde{\alpha} & \mathrm{if\:}\tilde{\alpha}>0,\\
-2\tilde{\alpha} & \mathrm{if\:}\tilde{\alpha}<0\,\wedge\,\textrm{guessInc}\left(u,z\right)=0,\\
\textrm{guessInc}\left(u,z\right)/2 & \mathrm{if\:}u\left(0\right)=1\,\vee\, u\left(0\right)=-1,\\
\textrm{guessInc}\left(u,z\right) & \mathrm{otherwise},\end{cases}\end{eqnarray*}

\vspace{-0.2in}
 \begin{eqnarray*}
\mbox{guessInc}\left(u,z\right)\;|\; u\:\mbox{is independent}\:\mbox{of}\: z & \rightarrow & 0.\end{eqnarray*}

The treatment of sums is more easily stated procedurally: Let $s$
be a sum and $\tilde{\sigma}=\mathrm{guessDE}(s,z)$. Let $\underline{S}$
be the set of all terms in $s$ that have $\tilde{\sigma}$ as the
guess for their dominant exponent, and let $\bar{S}$ be the set of
all the other terms. If $\bar{S}$ is empty, let $\tilde{\gamma}=0$.
Otherwise let $\tilde{\gamma}$ be the minimum guessed dominant exponent
of the terms in $\bar{S}$. Let $\tilde{\delta}=0$ if 0 is the guessed
increment for all of the terms in $\underline{S}$. Otherwise let
$\tilde{\delta}$ be the minimum non-zero guessed increment in $\underline{S}$.
Then the guessed increment for the sum is\[
\min\left(\tilde{\delta},\tilde{\gamma}-\tilde{\sigma}\right).\]

Because $\mathrm{guessDE}(\ldots)$ might return an incorrect guess
for a dominant exponent, it is possible for this $\mbox{guessInc}\left(\ldots\right)$
to return 0 when the increment is actually positive. Therefore we
can instead guess an increment of 1 if a \textsl{top-level} invocation
of $\mbox{guessInc}\left(\ldots\right)$ returns 0.

The required argument order in Table \ref{tab:necessaryOperandOrders}
for the inverse trigonometric and inverse hyperbolic functions depends
on whether the dominant exponent $\alpha$ of the argument $u$ is
positive, negative, or 0 with the corresponding coefficient being
a branch point. Evaluating $u\left(0\right)$ can help us decide this:
If $u\left(0\right)$ is a branch point, then we can use the above
iteration scheme on $u-u\left(0\right)$ to obtain the required order
$o\left(z^{k+\sigma}\right)$. If $u\left(0\right)\equiv0$, then
$\alpha>0$. If $u\left(0\right)$ is otherwise finite and non-zero,
then $\alpha=0$. Either way, the required order for $u$ is $o\left(z^{k}\right).$
Otherwise, either a negative dominant exponent or a coefficient having
a logarithmic singularity caused $u\left(0\right)$ to have infinite
magnitude or to be indeterminant. For such $u\left(0\right)$:\vspace{-0.1in}

\begin{itemize}
\item If $\mathrm{guessDE}\left(u,z\right)\geq0$, then we can request $o\left(z^{k}\right)$,
truncating if the resulting dominant exponent is actually negative.\vspace{-0.1in}

\item Otherwise we can request an iterative determination of the appropriate
order, with the proviso that the order be $o\left(z^{k}\right)$ if
$\alpha$ is actually positive, or else 0 with $c$ not a branch point.\vspace{-0.08in}

\end{itemize}
For a product of two or more operands, let $u$ contain a proper subset
of the operands and $v$ contain the complementary subset. Use recursion
if $u$ and/or $v$ is thereby a product. Letting $U$ and $V$ denote
the corresponding truncated series, we can guess initial dominant
exponents $\alpha$ and $\beta$, then alternatively increase them
if necessary until either $U$ or $V$ reveals its true dominant exponent.
Then we know enough to compute the other operand series to the appropriate
order without iteration, after which we know enough to truncate or
compute additional terms of its companion if necessary. We can terminate
and return 0 for both $U$ and $V$ if a guess for $\beta$ yields
$U=0+o\left(z^{k-\beta}\right)$, a guess for $\alpha$ yields $U=0+o\left(z^{k-\alpha}\right)$,
and $\alpha+\beta\geq k$. Algorithm 1 presents details.

\begin{algorithm}[ht]
\caption{Compute series($u$) and series($v$) for computing series($uv$)
to $o\left(z^{k}\right)$}

\textbf{Input}: Symbolic expressions $u$ and $v$, variable $z$,
and rational number $k$.\\
 \textbf{Output}: The ordered pair of truncated series {[}$U,V${]},
each computed to the necessary\\
 \vspace{0.05in}
 \hspace*{.75in}and sufficient order so that $UV$ is $o\left(z^{k}\right)$.\\
 $\delta_{u}\leftarrow-1;\quad\delta_{v}\leftarrow-1;\qquad$
// -1 means these increments haven't yet been computed\\
 $\alpha\leftarrow\mathrm{guessDE}\left(u,z\right);\quad\beta\leftarrow\mathrm{guessDE}\left(v,z\right)$;\\
 $m\leftarrow m_{0}\leftarrow k\!-\!\beta;\quad n\leftarrow n_{0}\leftarrow k\!-\!\alpha;$\\
 loop\\
 \hspace*{.2in}$U\leftarrow\mathrm{series}\left(u,\, z\!=\!0,\, o\left(z^{m}\right)\right);$\\
 \hspace*{.2in}if $U\neq0+o\left(z^{m}\right),$ then\\
 \hspace*{.4in  }$\alpha\leftarrow\mathrm{dominantExponent}\left(U\right);$\\
 \hspace*{.4in  }$n\leftarrow k-\alpha;$\\
 \hspace*{.4in  }$v\leftarrow\mathrm{series}\left(v,\, z\!=\!0,\, o\left(z^{n}\right)\right);$\\
 \hspace*{.4in  }if $V=0+o\left(z^{n}\right),$ then return $\left[U,V\right];$\\
 \hspace*{.4in  }$\beta\leftarrow\mathrm{dominantExponent}\left(V\right);$\\
 \hspace*{.4in  }if $m=k-\beta,$ then return $\left[U,V\right];$\\
 \hspace*{.4in  }if $m>k-\beta$, then return $\left[\mathrm{truncate}\left(U,k-\beta\right),V\right];$\\
 \hspace*{.4in  }return $\left[\mathrm{series}\left(u,\, z\!=\!0,\, o\left(z^{m-\beta}\right)\right),V\right]$;\\
 \hspace*{.2in}$V\leftarrow\mathrm{series}\left(v,\, z\!=\!0,\, o\left(z^{n}\right)\right);$\\
 \hspace*{.2in}if $V\neq0+o\left(z^{n}\right),$ then\\
 \hspace*{.4in  }$\beta\leftarrow\mathrm{dominantExponent}\left(V\right);$\\
 \hspace*{.4in  }$m\leftarrow k-\beta;$\\
 \hspace*{.4in  }$U\leftarrow\mathrm{series}\left(u,\, z\!=\!0,\, o\left(z^{m}\right)\right);$\\
 \hspace*{.4in  }if $U=0+o\left(z^{n}\right),$ then return $\left[U,V\right];$\\
 \hspace*{.4in  }$\alpha\leftarrow\mathrm{dominantExponent}\left(U\right);$\\
 \hspace*{.4in  }if $n=k-\alpha,$ then return $\left[U,V\right];$\\
 \hspace*{.4in  }if $n>k-\alpha,$ then return $\left[U,\mathrm{truncate}\left(V,k-\alpha\right)\right];$\\
 \hspace*{.4in  }return $\left[U,\mathrm{series}\left(v,\, z\!=\!0,\, o\left(z^{n-\alpha}\right)\right)\right];$\\
 \hspace*{.2in}if $m+n\geq k$, then return $\left[U,V\right];$\\
 \hspace*{.2in}$\delta_{u}\leftarrow\begin{cases}
1 & \delta_{u}<0\,\wedge\,\mathrm{guessInc}\left(u,z\right)=0,\\
\mathrm{guessInc}\left(u,z\right) & \delta_{u}<0,\\
\delta_{u}+\delta_{u} & \mathrm{otherwise};\end{cases}$\\
 \hspace*{.2in}$\delta_{v}\leftarrow\begin{cases}
1 & \delta_{v}<0\,\wedge\,\mathrm{guessInc}\left(v,z\right)=0,\\
\mathrm{guessInc}\left(v,z\right) & \delta_{v}<0,\\
\delta_{v}+\delta_{v} & \mathrm{otherwise};\end{cases}$\\
 \hspace*{.2in}$m\leftarrow\mathrm{min}\left(m_{0}+\delta_{u},k-n\right);$\\
 \hspace*{.2in}$n\leftarrow\mathrm{min}\left(n_{0}+\delta_{v},k-m\right);$\\
 endloop; 
\end{algorithm}

The chances for needing to truncate or iterate are reduced if we choose
for $v$ a factor for which the guess for $\beta$ is most likely
to be accurate, such as a linear combination of powers of $z$. To
aid this choice we can have $\textrm{guessDE}\left(\ldots\right)$
also return a status that is an element of\[
\left\{ lowerBound,\, exact,\, upperBound,\, uncertain\right\} ,\]
 with these constants being a guarantee about the guess. Example rules
for computing and propagating such a status are:\begin{eqnarray*}
\mathrm{statusOfGDE}\left(z,z\right) & \rightarrow & equal,\\
\mathrm{statusOfGDE}\left(u+v,z\right) & \rightarrow & lowerBound,\\
\mathrm{statusOfGDE}\left(u^{k},z\right) & \rightarrow & \begin{cases}
\mathrm{statusOfGDE}\left(u,z\right) & \mbox{if}\; k\geq1,\\
upperBound & \mbox{if}\;\mathrm{statusOfGDE}\left(u,z\right)=lowerBound,\\
lowerBound & \mbox{if}\;\mathrm{statusOfGDE}\left(u,z\right)=upperBound,\\
\mathrm{statusOfGDE}\left(u,z\right) & \mbox{otherwise.}\end{cases}\end{eqnarray*}

For example, if $\mathrm{statusAndGDE}\left(u,z\right)$ returned
$\left[lowerBound,6\right]$ and $\mathrm{statusAndGDE}\left(v\right)$
returned $\left[exact,4\right]$, then we know that $\mathrm{series}\left(uv,z=0,o\left(z^{9}\right)\right)$
is $0+o\left(z^{9}\right)$ without computing series for $u$ and
$v$.

We can return and exploit a similar status for $\textrm{guessInc}\left(\ldots\right)$.
For example, if\begin{eqnarray*}
\mathrm{statusAndGDE}\left(u,z\right) & \rightarrow & \left[lowerBound,6\right],\\
\mathrm{statusAndGInc}\left(u,z\right) & \rightarrow & \left[exact,0\right],\end{eqnarray*}
 then there is no point to iteratively increasing the requested order
for $u$ beyond 6.

Another way to compute operand series to the necessary and sufficient
order, pioneered by Norman \cite{Norman}, is to use lazy evaluation,
streams, or Lisp continuations. The idea is to generate at run time
a network of co-routines that recursively request additional order
or additional non-zero terms for sub-expressions on an incremental
as-needed basis. The above guesses and iterative techniques are relevant
there too, because if the request is for an increment to the order,
it might not produce another non-zero term and if the request is for
an additional non-zero term, iteration might be necessary to produce
it. However, for such algorithms that don't recompute all of the terms
each iteration, it is probably more efficient not to increase the
increment each iteration.

\subsection{Optional requested number of non-zero terms}

Often rather than a requested order, users need a requested number
of non-zero terms, regardless of what order is required to achieve
that. Most often the needed number of terms is 1. For example, a particularly
effective way to compute many limits is to compute the limit of the
dominant term. This is particularly helpful for indeterminacies of
the form $\infty-\infty$. As another example, if we equate a truncated
series to a constant, then it is much more likely that we can solve
this equation for $z$ if there are only one or two terms in the truncated
series. Thus it is important to implement a separate function such
as\[
\mbox{nTerms}\left(\textrm{\emph{expression}},\,\textrm{\emph{variable}}\!=\! point,\,\textrm{\emph{numberOfNonZeroTerms}}\right).\]

Parameter \textit{numberOfNonZeroTerms} could default to 1 and/or
there could be a separate function such as\[
\mbox{dominantTerm}\left(\textrm{\emph{expression}},\, variable\!=\! point\right).\]

For a hierarchical series the user often doesn't know \emph{a priori}
an appropriate set of basis functions for the series, and the dominant
basis function can be an essential singularity or a logarithm rather
than $z.$ For such series it is much more appropriate for users to
request the desired number of terms rather than an order in $z.$
In this context, it is appropriate to count recursively-displayed
terms of a hierarchical series as if they were fully expanded. For
example, only one such distributed term is necessary for purposes
such as computing a limit.

It is dangerously misleading to include a term unless all of the preceding
terms are fully developed, which might and often does require \textsl{infinite}
series for the \textsl{coefficients} of some preceding terms. For
example, it is inappropriate to include the $z^{3}$ term of\begin{equation}
z+\left(\sum_{k=0}^{\infty}\dfrac{\left(\ln z\right)^{k}}{2^{k+1}}\right)z^{2}+z^{3}+o\left(z^{3}\right)\label{eq:InfiniteSeriesCoefficient}\end{equation}
 if the series for the coefficient of $z^{2}$ is truncated, because
$\left(\ln z\right)^{k}z^{2}/2^{k+1}$ dominates $z^{3}$ for all
$k\geq0$. This is another reason that a requested distributed term
count is more appropriate than an order request for truncated hierarchical
series. This is also a good reason for providing the option of not
expanding coefficients as sub-series where the implementation can't
express them as infinite series and they don't terminate at a finite
number of terms. For example, there should be an option for even a
hierarchical series function to return\[
z+\left(\dfrac{1}{2-\ln z}\right)z^{2}+z^{3}+o\left(z^{3}\right)\]
 for expression (\ref{eq:InfiniteSeriesCoefficient})

To achieve a requested number of non-zero terms, we can iteratively
increase the requested order until we obtain at least that number
of non-zero terms, then truncate any excess terms. The iteration could
begin with a requested order somewhere in the interval $\mathrm{guessDE}\left(u,z\right)+\left[0,(n-1)\mathrm{guessInc}(u,z)\right]$,
where $n$ is the requested number of non-zero terms. If this attempt
exposes no terms, then we can increment the request by $n\cdot\Delta$
where $\Delta$ starts at $\mathrm{guessInc}(u,z)$ and doubles after
each failed attempt.

However, the implementation should address the fact that an expansion
might terminate as exact with fewer than the requested number of non-zero
terms. The fact that the returned number of terms is less than requested
is a subtle indication that the series is exact, but an explicit error-order
term of the form $\Theta\left(\left(\ldots\right)^{\infty}\right)$
makes this fact more noticeable. This is additional motivation for
having each intermediate series result include an indication of exactness,
if known, as elaborated in subsection \ref{sub:ComputationPropagationAndDisplayOfOrder}.

\section{Issues about displaying truncated series results}

In contrast to most other expressions, the terms of a truncated Puiseux
series expanded about $w=w_{0}$ for finite $w_{0}$ are traditionally
displayed in order of \textsl{increasing} powers of $w-w_{0}$, even
if there are logarithms involving $w$ in expressions multiplying
some of those powers. If series are represented using the same data
structures as general expressions but different ordering, then the
different ordering might prevent key cancellations because efficient
bottom-up simplification typically relies on the simplified operands
of every operator having the same canonical ordering. For example,\begin{eqnarray*}
-z+\mbox{series}\left(e^{z},\, z\!=\!0,\, o\left(z^{2}\right)\right) & \rightarrow & -z+\left(1+z+\dfrac{z^{2}}{2}\right)\\
 & \rightarrow & -z+1+z+\dfrac{z^{2}}{2},\end{eqnarray*}
 rather than simplifying all the way to $1+z^{2}/2$.

The $\mbox{series}\left(\ldots\right)$ function is most often used
alone as an input, perhaps with the result assigned to a variable,
rather than embedded in an expression. If so and the assigned value
is ordered normally or the system re-simplifies pasted and assigned
values when used in subsequent expressions, then the differently-ordered
series result would safely be re-ordered into the non-series order
during simplification of that subsequent input.

One way to overcome this difficulty entirely is to use a special data
structure for series results, then use a special method for displaying
those results. However, the next subsection describes how onerous
it is to integrate such special data structures into a system in a
thorough seamless way.

\subsection{The pros and cons of an explicit error order term.}

Maxima 5.18.1 displays {}``$+\ldots$'' at the end of a truncated
series result, which means $o\left(z^{n}\right)$ where $n$ is the
order argument provided by the user. Mathematica 7.0.1.0 displays
a big-$O$ term, and Maple 13.0 displays a big-$O$ term if the result
isn't exact. Even when result orders are always as requested, a displayed
ellipsis is useful and a displayed error term is even more valuable.
They remind users that although the result is symbolic, it is perhaps
or definitely approximate. Moreover, it provides an opportunity for
the implementation to make such truncated series \textsl{infectious},
which helps prevent users from misusing inappropriate mixtures of
approximate results with exact results or with results having different
orders or expansion points. For example, if a result of $\mbox{series}\left(\ldots\right)$
is\[
\left(w-2\pi\right)^{-\nicefrac{1}{2}}+\left(w-2\pi\right)^{2}+o\left(\left(w-2\pi\right)^{2}\right),\]
 then adding $\sin\left(w\right)$ to this result would return\[
\left(w-2\pi\right)^{-\nicefrac{1}{2}}+\left(w-2\pi\right)+\left(w-2\pi\right)^{2}+o\left(\left(w-2\pi\right)^{2}\right).\]
 With this infectiousness, $f\left(w\right)+o\left(\left(w-w_{0}\right)^{m}\right)$
is an elegant and convenient alternative to the input $\,\mbox{series}\left(f\left(w\right),\, w\!=w_{0},\, o\left(\left(w\!-\! w_{0}\right)^{m}\right)\right)$.

Unfortunately, the effort required to do a thorough job of implementing
this syntactic sugar is extensive. To correctly propagate the influence
of an error order term, every command, operator and function should
have a method for properly treating it. This obligation also applies
to every new command, operator or function that is subsequently added
to a system, including user-contributed ones that aspire to be first-class
citizens seamlessly integrated into the system.

Table \ref{tab:InfectiousErrorOrder} shows some examples that test
an implementation's handling of explicit error-order terms. Also,
test if the implementations you use can directly plot series results
or apply operators and functions such as $\int$, $\sum$, lim, $\textrm{solve}\left(\ldots\right)$,
and $\textrm{series}\left(\ldots\right)$ to series results without
the nuisance of first explicitly converting the series result to an
ordinary expression.

\begin{table}[ht]
\caption{Test examples for treating explicit error order terms\label{tab:InfectiousErrorOrder}}

\begin{tabular}{|c|c|}
\hline 
Input equivalent to  & Increasingly informative results\tabularnewline
\hline
\hline 
$z-\mbox{series}\left(z,\, z\!=\!0,\, o\left(z^{2}\right)\right)$  & $o\left(z\right)$, $O\left(z^{3}\right)$, 0\tabularnewline
\hline 
$\ln\left(\mbox{series}\left(e^{z},\, z\!=\!0,\, o\left(z^{2}\right)\right)\right)$  & $z\!+\! o\left(z^{2}\right)$, $z\!+\! O\left(z^{3}\right)$, $z\!+\!\Theta\left(z^{3}\right)$\tabularnewline
\hline 
$1+z-\mbox{series}\left(e^{z},\, z\!=\!0,\, o\left(z\right)\right)$  & $o\left(z\right)$, $O\left(z^{2}\right)$,$\Theta\left(z^{3}\right)$\tabularnewline
\hline 
$\mbox{series}\left(e^{z},\, z\!=\!0,\, o\left(z^{5}\right)\right)-\mbox{series}\left(e^{z}\, z\!=\!0,\, o\left(z^{2}\right)\right)$  & $o\left(z^{2}\right)$, $O\left(z^{3}\right)$, $\Theta\left(z^{3}\right)$\tabularnewline
\hline 
$\mbox{series}\left(e^{z}\!+\! z^{3},\, z\!=\!0,\, o\left(z^{2}\right)\right)-\mbox{series}\left(e^{z},\, z\!=\!0,\, o\left(z^{2}\right)\right)$  & $o\left(z^{2}\right)$, $O\left(z^{3}\right)$, $\Theta\left(z^{3}\right)$\tabularnewline
\hline 
$\dfrac{\mbox{series}\left(e^{z}\!+\! z^{3},\, z\!=\!0,\, o\left(z^{2}\right)\right)}{\mbox{series}\left(e^{z},\, z\!=\!0,\, o\left(z^{2}\right)\right)}$  & $1\!+\! o\left(z^{2}\right)$, $1\!+\! O\left(z^{3}\right)$, $1\!+\!\Theta\left(z^{3}\right)$\tabularnewline
\hline
\end{tabular}
\end{table}

\subsection{Computation, propagation and display of an order term\label{sub:ComputationPropagationAndDisplayOfOrder}}

With a little-$o$ interpretation of the $\mathrm{series}\left(\ldots\right)$
function order parameter and strict adherence to delivering as-requested
order, it is unnecessary to represent and propagate error-order during
the internal calculations, even if we display, $o\left(z^{\tau}\right)$
for that requested order.

Mathematica 7.0.1.0 and Maple 13.0 display an error order using $O$
rather than $o$. However, correctly determining a correct and satisfying
exponent to use in $O$ requires more work than $o$: To return a
result with a requested order $o\left(z^{\tau}\right)$ using a $O\left(z^{\nu}\right)$,
we must determine a $\nu>\tau$ such that the degree of the first
omitted non-zero term, if any, is at least $\nu.$ We can't just display
$O\left(z^{\tau+1}\right)$, because with fractional powers the exponent
of the first omitted term can be arbitrarily close to $\tau$.

One way to determine a correct $\tau$ is to actually compute the
first omitted non-zero term, but not display it. However, that omitted
term can have a degree arbitrarily greater than $\tau$, costing substantial
computation. Moreover, there might not be any non-zero terms having
degree greater than $\tau$. Therefore we don't know in advance what
order if any will just expose a next non-zero term whose coefficient
we discard. Also, if we find such a term, it would be more informative
to display $\Theta\left(z^{\nu}\right)$ rather than $O\left(z^{\nu}\right)$.

Another way to determine a correct $\nu$ is to compute a series that
is $o\left(z^{\tau+\Delta}\right)$ with a predetermined $\Delta$
such as 1, then truncate to $o\left(z^{\tau}\right)$ and display
$O\left(z^{\nu}\right)$ or $\Theta\left(z^{\nu}\right)$ where $\nu$
is the dominant degree of the truncated terms, or display $O\left(z^{\tau+\Delta}\right)$
if no terms were truncated. However, with fractional exponents there
can be arbitrarily many non-zero terms having exponents in the interval
$\left[\tau^{+},\tau+\Delta\right]$, which is costly. Moreover, users
might judge the implementation unfavorably if the exponent in $O$
is obviously less than it could be. For example with $\sin z$ the
series $z-z^{3}/3+O\left(z^{4}\right)$ is disturbing compared to
$z-z^{3}/3+O\left(z^{5}\right)$, which can be more informatively
displayed as $z-z^{3}/3+\Theta\left(z^{5}\right)$.

When computing series, we often know the exact order for the series
of some or all sub-expressions. For example, if the requested order
is $3$ then $z^{2}$ is $\Theta\left(z^{\infty}\right)$, whereas
$z^{5}$ is $\Theta\left(z^{5}\right)$. If we decide to store error-order
information with the series for each sub-expression, then it preserves
information to store with the error order whether it is of type $o$,
$O$, or $\Theta,$ and to propagate it according to rules such as,
for $\alpha<\beta$;\begin{eqnarray*}
\Theta\left(z^{\alpha}\right)+\Theta\left(z^{\beta}\right) & \rightarrow & \Theta\left(z^{\alpha}\right),\\
\Theta\left(z^{\alpha}\right)+\Theta\left(z^{\alpha}\right) & \rightarrow & O\left(z^{\alpha}\right),\\
\Theta\left(z^{\alpha}\right)+O\left(z^{\alpha}\right) & \rightarrow & O\left(z^{\alpha}\right),\\
\Theta\left(z^{\alpha}\right)+o\left(z^{\alpha}\right) & \rightarrow & \Theta\left(z^{\alpha}\right),\\
o\left(z^{\alpha}\right)+O\left(z^{\alpha}\right) & \rightarrow & O\left(z^{\alpha}\right).\end{eqnarray*}

\section{\label{sec:representation}A frugal dense representation}

A sparse series representation can more generally accommodate truncated
Hahn series, which can also have \textsl{irrational} real exponents.
For example,\[
z^{\sqrt{2}}+z^{\pi}+z^{4}+\ldots\]
 This extra generality is desirable for hierarchical series. Adaptive-precision
interval arithmetic can be used to keep the exponents properly ordered.

However, most published algorithms for series are written in a notation
that encourages a dense representation as an array or list of coefficients
with implied exponents. Adding two series is easy for sparse representation.
Otherwise, adapting most of the published truncated power series algorithms
to a sparse representation seems likely to make them more complicated.
Moreover, with typical applications dense representation is efficient
for most univariate polynomials, hence also for most series. As described
in \cite{StoutemyerPolyRep} recursive dense representation is also
surprisingly efficient for most sparse multivariate polynomials, hence
also for recursive hierarchical series or multi-variate series. Therefore
this section describes a particularly efficient dense representation
and some algorithmic necessities for maintaining it.

The allowance of negative exponents suggests that we should also explicitly
store the exponent of the dominant term.

The allowance of fractional exponents suggests that we should also
store the implicit positive rational exponent increment between successive
stored coefficients. To minimize the number of stored 0 coefficients,
it is most efficient to make this increment be the greatest common
divisor of all the exponent increments between successive non-zero
coefficients. (The gcd of two reduced fractions is the gcd of their
numerators divided by the least common multiple of their denominators).

The truncated series 0 can be represented canonically as a leading
exponent of 0, an exponent increment of 0, and consistently either
no coefficients or one coefficient that is 0. Using no coefficients
is more frugal and easier to program, but one zero coefficient permits
distinguishing a floating-point series $0.0+o\left(z^{\tau}\right)$
from a rational-coefficient series $0+o\left(z^{\tau}\right)$.

Rather than the gcd of the exponent increments, many implementation
instead use the reciprocal of the \textsl{common denominator} of all
the exponents. However, this can require arbitrarily more space and
time. For example, it would store and process 21 coefficients rather
than 3 for $1+2z^{10/3}+3z^{20/3},$ and it would store and process
31 coefficients rather than 4 for $z^{-10}+2+3z^{20}$.

Either way, for canonicality, programming safety and efficiency, it
is important to trim leading, trailing and excessive intermediate
zero coefficients from intermediate and final results wherever practical.
However, \textsl{within} a function that adds two series, etc., it
might be convenient to temporarily use series that have leading 0
coefficients and/or an exponent increment that is larger than necessary.
For example: 
\begin{itemize}
\item When two series having different exponent increments are \textsl{multiplied},
we can use copies in which extra zeros are inserted between the given
coefficients of one or both series so that their mutual exponent increment
is the gcd of the two series exponent increments. 
\item Let $\gamma$ be the gcd of the dominant exponents and exponent increments
of two series. If the series have different dominant exponents and/or
different exponent increments, then before the series are \textsl{added},
copies of one or both series can be padded with extra zeros before
the dominant coefficient and/or between coefficients so that both
series start with the same implicit exponent and have the same implicit
exponent increment $\gamma$. 
\end{itemize}
For both examples the resulting series should then be adjusted if
necessary so that its leading coefficient is non-zero and its exponent
increment is as large as possible.

When computing any one series, the sub-expressions all have the same
expansion variable and expansion point 0 after the transformations
described in Section 1. Also, the desired order of the result is specified
by the user and can be passed into the recursive calls for sub-expressions,
adjusted according to Table \ref{tab:InfectiousErrorOrder}. If an
implementation delivers an as-requested $o$-order, then there is
no need to store it in the series data structure for intermediate
series results. Therefore, only the dominant exponent, exponent increment
and \textsl{frugalized} coefficient list or array are necessary for
an \textsl{internal} data structure \textsl{during} computation of
any one series.

For each function or operator, such as $\ln$ and {}``+'', it is
helpful to have a function that, given an expression and a requested
order for expansion at $z=0$:\vspace{-0.1in}

\begin{enumerate}
\item guesses the dominant exponents of the operand series where needed,\vspace{-0.1in}

\item computes the guessed necessary and sufficient order to request for
the operand series from Table \ref{tab:InfectiousErrorOrder} and
the guessed dominant exponents,\vspace{-0.1in}

\item recursively computes those series to the guessed necessary and sufficient
orders,\vspace{-0.1in}

\item truncates if the requested orders are excessive, or iteratively increases
the requested orders if they are insufficient,\vspace{-0.1in}

\item invokes a companion lower-level function to compute the result series
from the resulting operand series. (For computing a function of a
given \textsl{series}, this companion function would be invoked directly.) 
\end{enumerate}
If we wish to report to the user the type of the resulting order ($\theta$,
$o$, or $O$) and the corresponding exponent, then the internal data
structure must also contain fields for those.

If we also wish to preserve with the final result the expansion variable
and expansion point, then we must have an \textsl{external} data structure
that includes those together with the internal data structure.

\section{Exponentials interact with logarithmic coefficients}
\begin{defn*}
A function $f\left(z\right)$ is \textsl{sub-polynomial }with respect
to $z$ if \begin{eqnarray*}
\lim_{r\rightarrow0^{+}}\,\left|\left(re^{i\theta}\right)^{\gamma}f\left(re^{i\theta}\right)\right| & = & \begin{cases}
0 & \forall\gamma>0,\\
\infty & \forall\gamma<0.\end{cases}\end{eqnarray*}

\end{defn*}
Examples include\vspace{-0.1in}

\begin{itemize}
\item an expression independent of the expansion variable $z$, or\vspace{-0.1in}

\item an expression that is piecewise constant with respect to $z$, or\vspace{-0.1in}

\item an expression of the form $\ln\left(c\left(z\right)z^{\alpha}\right)$,
where $\alpha\in\mathbb{Q}$ and $c\left(z\right)$ is sub-polynomial,
or \vspace{-0.1in}

\item any sub-exponential function of sub-polynomial expressions.\end{itemize}
\begin{defn*}
A sub-exponential function $g\left(z\right)$ is one for which $g\left(\ln z\right)$
is sub-polynomial.
\end{defn*}
Examples of sub-exponential functions include rational functions,
fractional powers, $\ln$, inverse trigonometric and inverse hyperbolic
functions.

Non-constant sub-polynomial coefficients can arise for a series of
an expression that contains logarithms, inverse trigonometric or inverse
hyperbolic functions of the series variable.

Most Puiseux-series algorithms require no change for coefficients
that are generalized from constants to sub-polynomial expressions,
making this powerful generalization of Puiseux series cost very little
additional program space. For example, some algorithms for computing
functions of \textsl{constant-coefficient} Taylor series are derived
via a differential equation. However, once the algorithms are obtained
for constant coefficients, there is no need to incur the difficulties
of including any dependent coefficients in the differentiations and
integrations used in the derivations.

However, the algorithm for computing the exponential of a series does
require a change: For series having a non-negative dominant exponent,
the algorithm begins by computing the exponential of the degree-0
term to use in the result coefficients. If the degree-0 term is a
multiple of a logarithm of a monomial containing a power of $z$,
then the exponential of the leading coefficient generates a power
of $z$. To avoid incorrect truncation levels, this power of $z$
should be combined with the (perhaps implicit) power portions of the
data structure so that the true degree of each term is manifest in
a canonical way.

Other places where coefficients interact with powers are computing
derivatives or integrals of a series with respect to $z$. For example,
$\frac{d}{dz}\ln z\rightarrow z^{-1}$, and $\int\ln z\, dz\rightarrow\left(\ln z-1\right)z$.

\section{Essential singularities}

Exponentials and sinusoids of negative powers of $z$ are essential
singularities at $z=0$. For a series, let $\underline{U}$ be the
sum of the terms having negative exponents and $\overline{U}$ be
the sum of all the other terms. We could use the transformation $e^{\underline{U}+\overline{U}}\rightarrow e^{\underline{U}}\cdot e^{\overline{U}},$
then compute the series for $e^{\overline{U}}$, then distribute the
essential singularity $e^{\underline{U}}$ over the resulting terms
as factors in the coefficients. We could similarly use angle sum transformations
for sinusoids of series having negative exponents. However, essential
singularities dominate any power of $z$ at $z=0$. If we distribute
the essential singularity, then subsequent series operations can truncate
terms that dominate retained terms that don't contain the essential
singularity, giving an incorrect result. For example,\begin{eqnarray*}
\mathrm{series}\left(e^{z^{-1}}\left(1+z^{2}\right)+\sin z,\, z=0,\, o(z)\right) & \longrightarrow & \left(e^{z^{-1}}+e^{z^{-1}}z^{2}\right)+\left(z+o(z)\right)\\
 & \stackrel{incorrect}{\longrightarrow} & e^{z^{-1}}+z+o(z).\end{eqnarray*}

Therefore it is more appropriate to produce a recursively represented
series in $e^{\underline{U}}$ having coefficients that are generalized
Puiseux series in $z$ -- a hierarchical series.

We could have one extra field in our data structure for a multiplicative
essential singularity. However, that complicates the algorithms for
very little gain, because subsequent operations can easily require
a more general representation. For example, one field for a single
multiplicative essential singularity can't represent \[
e^{z^{-2}}+z.\]
 The extra effort of implementing such a limited ability to handle
essential singularities is better spent implementing more general
hierarchical series.

Collecting exponentials in an expression permits computation of generalized
Puiseux series for some expressions containing essential singularities
that are canceled by the collection. For example,\begin{eqnarray*}
\dfrac{e^{\csc z}}{e^{\cot z}} & \rightarrow & e^{\csc z-\cot z}\rightarrow e^{z/2+z^{3}/24+\cdots}\rightarrow1+\dfrac{z}{2}+\dfrac{z^{2}}{8}+\cdots,\\
\left(e^{1/x}\right)^{\sin x}\,|x\in\mathbb{R} & \rightarrow & e^{\left(\sin x\right)/x}\rightarrow e^{1-x^{2}/6+\cdots}\rightarrow e-\dfrac{ex^{2}}{6}+\cdots.\end{eqnarray*}

\section{Unnecessary Restrictions}

Not all Puiseux-series implementations currently allow fractional
requested order. However, if fractional exponents are allowed in the
result, then it is important to permit them as the requested order
too. Otherwise, for example, a user will have to compute and view
1001 terms of $\exp\left(z^{1/1000}\right)$ merely to see the first
two terms $1+z^{1/1000}$.

Not all implementations currently allow negative requested order.
However, if negative exponents are allowed in the result, then it
is important to permit them as the requested order too. Otherwise,
for example, a user will have to compute and view 1001 terms of $e^{z}/z^{1000}$
merely to see the first two terms $z^{-1000}+z^{-999}$.

These restrictions are probably caused by restricting some field in
the data structure to a one-word signed or unsigned integer, which
can also unnecessarily limit the magnitude of the requested order.
Although most likely motivated by a desire for efficiency, the savings
are probably a negligible percentage of the time consumed by coefficient
operations and other tasks.

Not all implementations currently allow non-real infinite-magnitude
expansion points, such as for\begin{eqnarray*}
\mathrm{series}\left(w^{-1},\, w\!=\! i\infty,\,3\right) & \rightarrow & w^{-1}+\theta\left(w^{\infty}\right),\end{eqnarray*}
 despite the fact that such limit points can be mapped to a real infinity
by a transformation such as $w\rightarrow-iz$.

Not all implementations currently allow full generality for sub-polynomial
coefficients. For example,\begin{eqnarray*}
\mathrm{series}\left(\arcsin\left(\ln z\right),\, z\!=\!0,\,3\right) & \rightarrow & \arcsin\left(\ln z\right)+\theta\left(z^{\infty}\right).\end{eqnarray*}
 The sub-polynomial coefficient of $z^{0}$ can be developed as a
truncated hierarchical infinite series $\ln z+(\ln z)^{3}/6+\cdots$,
which is preferable for some purposes such as computing a limit at
$z=0$. However, if the request is for expansion in powers of $z$.
then $\arcsin\left(\ln z\right)$ has the advantage of being exact
and much simpler.

\section*{Summary}

The generalization from Taylor series to generalized Puiseux series
introduces a surprising number of difficulties that haven't been fully
addressed in previous literature and implementations. Such issues
discussed in this article are:\vspace{-0.1in}

\begin{itemize}
\item avoiding unnecessary restrictions such as prohibiting negative or
fractional orders,\vspace{-0.1in}

\item the pros and cons of displaying results with explicit infectious error
terms of the form $o\left(\ldots\right)$, $O\left(\ldots\right)$,
and/or $\Theta\left(\ldots\right)$,\vspace{-0.1in}

\item efficient data structures, and\vspace{-0.1in}

\item algorithms that efficiently give users exactly the order they request. 
\end{itemize}

\section*{Acknowledgments}

I thank David Diminnie and Arthur Norman for their assistance.

\end{document}